\documentstyle[11pt,a4,epsfig]{article}
\newcommand{\micron}{$\mu{\rm m}$}
\newcommand{\chd}{${\rm CH_3D}$}
\newcommand{\ch}{${\rm CH_4}$}
\newcommand{\h}{${\rm H_2}$}
\newcommand{\ho}{${\rm H_2 O}$}
\begin{document}

\begin{center}
{\Large\bf Vapor pressure isotope fractionation effects in planetary atmospheres: application to deuterium.}\\[2cm]

Thierry Fouchet and Emmanuel Lellouch\\[2cm]

DESPA, Observatoire de Paris, 5 Place Jules Janssen, 92195 Meudon Cedex, France\\
tel.: 33.1.45.07.76.60\\
fax : 33.1.45.07.71.10\\
email : Thierry.Fouchet@obspm.fr\\[8cm]

24 pages\\
1 table\\
6 figures\\
\end{center}

\newpage
\noindent Proposed running title: Vapor pressure isotope effect in planetary atmospheres\\

\noindent Send correspondence to: Thierry Fouchet, Observatoire de Paris, DESPA-T1m, 5 Place Jules Janssen, 92195 Meudon Cedex, France; tel.: 33.1.45.07.76.60; fax: 33.1.45.07.71.10; email: Thierry.Fouchet@obspm.fr
\newpage
\section*{Abstract}

\indent\indent The impact of the vapor pressure difference between deuterated and nondeuterated condensing molecules in planetary atmospheres is quantitatively assessed. This difference results in a loss of deuterium in the vapor phase above the condensation level. In Titan, Uranus and Neptune, the effect on \chd\ is too subtle to alter current D/H ratio determinations. In Mars, the effect can induce a large depletion of HDO, starting about one scale height above the condensation level. Although the current infrared measurements of the D/H ratio appear to be almost unaffected, the intensity of disk-averaged millimetric HDO lines can be modified by about 10\%. The effect is much stronger in limb sounding, and can be easily detected from orbiter observations.\\

{\em Key words:} Atmospheres, Composition; Mars, Atmosphere; Isotopes

\newpage
\section{Introduction}

\indent\indent Since the first detection of \chd\ in the atmosphere of Jupiter (Beer and Taylor 1973), the D/H ratio, and to a lesser extent other isotopic ratios, have played a major role in the study of the origin and evolution of planetary atmospheres. Indeed, the large mass difference between H and its isotope D induces large differences in both the thermodynamics and kinetics of various processes, such as thermal escape, chemical reactions and condensation, resulting in an isotopic fractionation. In turn, the observed D/H ratios act as tracers of the chain of events which have formed the currently observed planetary atmospheres from the initial interstellar ices and gases. For this reason many observational efforts have focussed on the precise measurement of the D/H ratios in Solar System objects (see Owen (1992) for a review).

However, in many circumstances, the D/H ratio observed in a particular molecular species does not represent the bulk composition of the planet, if any fractionation occurs at the sounded atmospheric levels. This is the case in the Giant Planets, where the D/H observed in methane (i.e.\ from \chd) generally differs from that observed in hydrogen (i.e.\ HD), because isotopic exchange occurs in chemical reactions between \ch\ and \h\ (Beer and Taylor 1973, Fegley and Prinn 1988, Lecluse {\it et al.}\ 1996). This effect has been extensively studied and estimates of the corresponding fractionation factor $f$ are available, though still uncertain.

Another possible, but much less studied effect, is the isotopic fractionation caused by condensation, which can take place if the D/H ratio is observed in a condensible species, such as \ho\ in Mars, or \ch\ in Titan, Uranus and Neptune. This effect arises from some slight differences between the vapor pressures of isotopes of a given chemical species, with the heavy isotope molecule generally having a lower vapor pressure. During cloud formation in a planetary atmosphere, the condensed phase is then enriched, and the vapor phase depleted, in the heavy isotope. The depletion increases with the amount of condensed mass. This effect is widely documented on Earth, both theoretically and observationally, with aircraft measurements determining a D/H ratio as low as 0.7 times the standard ratio (SMOW) a few kilometers above the condensation level (Taylor 1972). This raises concern that the D/H ratios measured in other planetary atmospheres could also suffer from significant departures from the global planetary ratios, due to condensation fractionation. 

The subject of this paper is to investigate the effect of the isotopic fractionation occuring in cloud condensation, and how it affects the measured D/H ratio. We focus on water in the atmosphere of Mars, and on methane in Titan, Uranus and Neptune. In the next section we present the theoretical bases of isotopic variations induced by cloud formation, along with the available laboratory data relevant to the studied planets. Then we present the expected isotopic variations and the consequences on the observed D/H ratios. In the case of Mars, we also present synthetic calculations of millimetric HDO lines in order to determine the detectability of this isotopic fractionation.

\section{Laboratory measurements and isotopic cloud models}

\subsection{Fractionation coefficient}

\indent\indent The difference in equilibrium vapor pressures of isotopic molecules, often referred to as the vapor pressure isotope effect (VPIE), exhibits a large variety of behaviors, revealing the nature of the intramolecular forces at work in the condensed phases. For this reason, many laboratory and theoretical studies have been aimed at precisely measuring or modeling the VPIE, in particular for water and methane. A comparison of the different values can be found in Jancso and Van Hook (1974), but here, following Jouzel (1986) and Bigeleisen {\it et al.}\ (1967), we adopt the fractionation coefficients of Merlivat and Nief (1967) for water and of Armstrong {\it et al.}\ (1953) for methane.

The isotopic content of an atmospheric parcel can be expressed in terms of $\delta$:
\begin{equation}
\delta={({\rm D/H})_{parcel}-({\rm D/H})_{\circ}\over ({\rm D/H})_{\circ}},
\label{}
\end{equation}
where $({\rm D/H})_{\circ}$ is the standard atmospheric value, i.e.\ the global D/H ratio of the planet. The VPIE is expressed in terms of a fractionation coefficient $\alpha$, defined as the ratio of the D/H in the condensed phase to that in the vapor phase:
\begin{equation}
\alpha={({\rm D/H})_c\over ({\rm D/H})_v}={1+\delta_c\over 1+\delta_v},
\label{Eq:0}
\end{equation}
where the $c$ and $v$ subscripts refer respectively to the condensed and vapor phases.

The water fractionation coefficient for the solid/vapor transition is given by Merlivat and Nief (1967):
\begin{equation}
\ln\alpha = {16.288\over T^2}\times10^3-9.34\times10^{-2},
\label{}
\end{equation}
where $T$ is the temperature in Kelvin. Unfortunately, this expression is experimentally validated only between 230 and 360\,K. No data are available at the temperatures relevant for Mars' atmosphere, where water condensation occurs typically between 190\,K and 140\,K. The extrapolation of the above formula down to these temperatures might be hazardous. Therefore, we explore the consequences of isotopic fractionation in two cases. In the first one, we extrapolate the expression of Merlivat and Nief (1967) to the full Martian temperature range ($\alpha=1.43$ at 190\,K, and $\alpha=2.09$ at 140\,K). In a second, conservative approach, we adopt the coefficient at 230\,K ($\alpha=1.24$) for all lower temperatures. This likely underestimates the effects of isotopic fractionation.

The methane fractionation coefficient has been measured for the solid/vapor transition by Armstrong {\it et al.}\ (1953):
\begin{equation}
\ln\alpha = {110.2\over T^2}-{1.26\over T},
\label{}
\end{equation}
where $T$ is the temperature in Kelvin. This expression is based on laboratory measurements carried out between 75 and 91\,K. The temperature minimum in Uranus and Neptune lies around 55\,K. However at this temperature the extrapolated fractionation coefficient is still very low ($\alpha=1.014$). In addition, Bigeleisen {\it et al.}\ (1967) gave a theoretical justification to the expression of Armstrong {\it et al.} (1953). Therefore, we think it can safely be extrapolated down to 55\,K. Methane exhibits the rare behavior of an inverse isotope effect: the condensed phase is enriched in the lightest isotope (i.e.\ \ch) at temperatures warmer than 87.5\,K, while at lower temperatures the ``normal'' effect (i.e.\ enrichment in \chd), takes place. However, as in Titan's, Uranus' and Neptune's atmospheres methane condenses only at lower temperatures, the global effect is a depletion of deuterium in the gas phase, as we show below.

\subsection{Cloud models}

\indent\indent Cloud formation occurs when a parcel of moist air cools below its dewpoint. Cooling can be due to a variety of processes: radiative heat loss, mixing of different air masses, expansion due to vertical lifting. In the planetary atmospheres considered here, most of cloud condensation is caused by vertical lifting, either in convection for Uranus and Neptune (Carlson {\it et al.}\ 1988), or in global circulation Hadley cells on Mars (Jakosky and Haberle 1992) and Titan (Hourdin {\it et al.}\ 1995). For this type of cooling, the isotopic fractionation theory has been extensively developed for the Earth's atmosphere (see Jouzel (1986) for a review). Modeling of actual clouds is difficult, due to the fact that part of the condensed phase is removed by precipitation, while the remaining part ascends with the updraft. Moreover, in planetary atmospheres the condensation processes are still poorly understood, compared to the Earth's atmosphere. Therefore we restrict our study to two simple ideal cases considered by Dansgaard (1964) and designed to span the whole range of possible cloud models, and thus of isotopic fractionation values.

In the following, we assume that the condensed and vapor phases are always in isotopic equilibrium, i.e.\ that the isotopic equilibrium is kinetically achieved. This is clearly not the case in the Earth's atmosphere (Taylor 1972). Jouzel {\it et al.}\ (1980) have shown that neglecting kinetics effects, in the case of hailstone formation in the Earth's atmosphere, leads to a 20\% overestimation of the D/H depletion. However, modeling this departure is beyond the scope of this work which aims at a first order estimate of the VPIE in planetary atmospheres.

The first model is based on an open cloud system (Dansgaard 1964), where the solid phase is assumed to condense in isotopic equilibrium with the vapor phase, and to leave the cloud by precipitation immediately after its formation. In this case, the elementary isotopic variation of the vapor phase is given by:
\begin{equation}
{\rm d}(1+\delta_v)={{\rm d}({\rm D/H})_v\over ({\rm D/H})_{\circ}}={1\over ({\rm D/H})_{\circ}}\left(\left({n^i_v+{\rm d}n^i_v\over n_v+{\rm d}n_v}\right)-{n^i_v\over n_v}\right),
\label{}
\end{equation}
leading at first order to:
\begin{equation}
{\rm d}(1+\delta_v)={1\over ({\rm D/H})_{\circ}}{n^i_v\over n_v}\left({{\rm d}n^i_v\over n^i_v}-{{\rm d}n_v\over n_v}\right),
\label{}
\end{equation}
where $n_v$ and $n^i_v$ are the vapor mixing ratios respectively of the nondeuterated and deuterated species, and ${\rm d}n_v$ and ${\rm d}n^i_v$ their elementary variations. Since the condensed phase is in isotopic equilibrium with the vapor phase we have the relation:
\begin{equation}
{{\rm d}n^i_v\over {\rm d}n_v}=\alpha{n^i_v\over n_v},
\label{}
\end{equation}
from which it results that:
\begin{equation}
{{\rm d}\delta_v\over 1+\delta_v}=(\alpha-1){{\rm d}n_v\over n_v}.
\label{eq:1}
\end{equation}
The open cloud system gives the largest possible depletion of the gas phase in deuterium above the condensation level, since the isotopic excess in the condensed phase is definitely and constantly removed by precipitation.

In contrast, the second model induces the smallest possible depletion. It is based on a closed cloud system (Dansgaard 1964), where the condensed and vapor phases ascend together in the air parcel. In this case the number of molecules is conserved:
\begin{equation}
\left\{n_{\circ}=n_c+n_v\qquad\qquad\hbox{for the nondeuterated species}\atop n_{\circ}\delta_{\circ}=n_v\delta_v+n_c\delta_c\hfill\hbox{for the deuterated species}\right.
\label{eq:cons}
\end{equation}
where $n_v$ (respectively $n_c$) is the gas mixing ratio of the nondeuterated species in the vapor phase (respectively condensed) phase, and $n_{\circ}$ and $\delta_{\circ}$ are the mixing ratio of the nondeuterated species and the isotope ratio below the condensation level. It thus results from the combination of Eq.~\ref{Eq:0} (isotopic equilibrium) and Eq.~\ref{eq:cons}, that the isotopic ratio of the vapor phase is given by:
\begin{equation}
\delta_v={n_{\circ}(\delta_{\circ}-\alpha+1)+(\alpha-1)n_v\over(1-\alpha)n_v+\alpha n_{\circ}}.
\label{eq:2}
\end{equation}
  
These two theoretical models have been experimentally validated for the terrestrial atmosphere. Taylor (1972) has shown from in-situ measurements that the deuterium content of air masses above the condensation level was in agreement with the open cloud model, while laboratory analyses of hailstones (Jouzel {\it et al.}, 1975) support the closed cloud model under stormy conditions. In addition, we have validated our code by running it in terrestrial conditions, and comparing our results with the calculations of Jouzel (1986) (their Fig.~4).

\section{Atmospheric modeling}

\subsection{Mars: HDO}

\indent\indent The Martian D/H ratio was determined in the infrared by Owen {\it et al.}\ (1988) and Krasnopolsky {\it et al.}\ (1997) at 3.7\,\micron, and by Bjoraker {\it et al.}\ (1989) at 2.65\,\micron. In this solar reflected part of the spectrum, the observations are sensitive to the whole column density, and the D/H ratio is estimated by comparing the total column densities of \ho\ and HDO. For this reason, the impact of the VPIE on the D/H determination strongly varies with the altitude of water condensation. If condensation occurs close to the surface a large fraction of the HDO column can be affected by the isotopic fractionation. If it occurs a few scale heights above the surface the modification to the column is only marginal. A proper estimate of the VPIE on Mars must then take into account the variations of the water condensation level, due to variations in temperatures and water content of the lower atmosphere.

The Martian climate exhibits strong annual variations due to the high eccentricity of the planet's orbit, on which are superimposed episodes of global warming correlated with the dust content of the atmosphere (Clancy {\it et al.}\ 1996). We investigate the effect of the isotopic fractionation in the case of three different temperature profiles. In the lower atmosphere, we select profiles from the disk-averaged measurements of Clancy {\it et al.}\ (1996). Such observations are most appropriate to model the effects of isotopic fractionation as they would be observed from ground based observations of the planet. The millimeter wave observations of Clancy {\it et al.}\ (1996) typically constrain the atmospheric temperature up to ~45\,km (0.05\,mbar). At higher levels, we take temperature information from the numerical simulations of the Martian Climate Database (MCD) (Read {\it et al.}\ 1997, {\it http://www.lmd.jussieu.fr/mars.html}). Our three thermal profiles are the following:
\begin{itemize}
\item {\bf Cold profile:} at Mars' aphelion, in the case of a low dust loading of the atmosphere, we adopt the coldest temperatures of Clancy {\it et al.}\ (1996): 190\,K at 3\,mbar, 160\,K at 0.5\,mbar and 140\,K at 0.05\,mbar (Fig.~1a). Following the MCD, the temperature is left constant at 140\,K above this pressure level.
\item {\bf Mean profile:} at Mars' perihelion, in the case of a low dust loading of the atmosphere, the temperatures are 210\,K at 3\,mbar and 180\,K at 0.5\,mbar (Fig.~2a). Following the MCD, we place the tropopause at 0.05\,mbar with a temperature of 140\,K.
\item {\bf Hot profile:} at Mars' perihelion and with a high dust loading, we adopt a temperature profile in agreement with the April 1992 measurement of Clancy {\it et al.}\ (1996) and with that observed by Viking/IRTM (Martin and Kieffer 1979). This profile has  220\,K at 3\,mbar, 200\,K at 0.5\,mbar, and 180\,K at 0.05\,mbar. The tropopause is fixed at 5\,$\mu$bar and 140\,K.
\end{itemize}

The amount of water in the Martian atmosphere is highly variable, both spatially and seasonally (Jakosky and Haberle 1992), as a result of exchanges between the atmosphere and non-atmospheric reservoirs, such as polar caps or ground ice. In order to investigate the largest set of possible situations we calculate the isotopic fractionation for each of the three temperature profiles with three typical water column densities: 3, 10 and 30 precipitable micrometers (pr.\,\micron).

Although atmospheric water vapor is in equilibrium with the ice reservoirs, the global (i.e.\ below the clouds) atmospheric D/H ratio is not expected to differ from that in water ice. This is because fractionation effects occur at each phase change except at sublimation and melting of ice (Jouzel 1986). This behavior comes from the very slow molecular diffusion within ice, which makes the time scale of isotope exchanges inside the ice longer than the time scale of phase change (Joussaume {\it et al.}, 1984). (For the same reason, the D/H ratio measured in cometary comae should represent the true nuclear value.)

For the three different temperature profiles and water column densities we calculate the expected vertical profiles of the D/H ratio (more precisely of $\delta_v$). The results are displayed on Fig.~1, Fig.~2 and Fig.~3. In the modeling, we assume that the condensation process occurs at 100\% relative humidity, i.e.\ we do not allow for supersaturation. The closed cloud model gives rise to a moderate depletion in HDO. $\delta_v$ asymptotically converges toward $-19\%$ at the top of the atmosphere if $\alpha$ is kept constant to its value at 230\,K. In the temperature-dependent case the depletion is larger: $\delta_v=-52\%$. These values can be inferred from the asymptotic value of Eq.~\ref{eq:2}: $\delta_v=(1-\alpha)/\alpha$, where $\alpha$ is taken at 230\,K and 140\,K respectively. In contrast, the open cloud model induces a huge depletion in HDO in both cases. The D/H ratio at the tropopause is $\sim20\%$  of its value at the surface ($\delta_v\sim-80\%$) if $\alpha$ is kept constant, and only $\sim1\%$ ($\delta_v\sim-99\%$) in the temperature-dependent case.

To estimate the effect of the isotopic fractionation on the D/H determinations, we compare the HDO column density from the models with that expected for an altitude independent D/H ratio. The results of this calculation are given in Table~I. As expected, the column integrated HDO depletion strongly depends on the level of cloud formation, and consequently on the temperature profile. The hot and mean profiles lead to a very small depletion, less than 1\% (Table~I). On the contrary, a depletion of the HDO column of the order of 10\% occurs for the cold profile, i.e.\ when the condensation level lies close to the surface. Note that the depletion in the HDO column is less sensitive to the \ho\ amount than to the temperature profile.

Because of the uncertainty on the behavior of real martian clouds and on the extrapolation of the fractionation coefficient, it seems difficult to accurately correct for the condensation effects. However, in what follows, we examine to which extent past and future observations of HDO/\ho\ may be affected.

Measurements of the Martian D/H ratio have been reported by Owen {\it et al.}\ (1988) and Krasnopolsky {\it et al.}\ (1997). However, their observational uncertainties, ${\rm D/H}=6\pm3$ and $5.5\pm2$ times the standard terrestrial ratio (SMOW), are so large that taking into account the fractionation effect has little value. Bjoraker {\it et al.}\ (1989) obtained observations of Mars near perihelion (August 5, 1988), and reported a much more precise value of $5.2\pm0.2$ times the standard terrestrial ratio. Since these data have not been published extensively, it is difficult to assess their accuracy. In any case, at the time of these observations, the observed water abundance was close to 10 pr.\,\micron, and the temperature measured at 5.1\,mbar was high ($228$\,K). Our hot profile is thus the most appropriate to estimate the HDO column depletion, showing that the reported D/H ratio is essentially unaltered by the fractionation effect.

HDO has been also detected in the microwave range (Encrenaz {\it et al.}\ 1991, Clancy {\it et al.}\ 1992), and used as an estimate of \ho\ assuming the D/H ratio of Bjoraker {\it et al.}\ (1989). The vapor pressure isotope effect could bias the water abundance determination. Comparison of HDO and \ho\ or ${\rm H_2{}^{18}O}$ observations in this spectral range could also offer an opportunity to detect this effect. This is shown in Fig.~4, where we compare synthetic profiles of the 226\,GHz HDO line, calculated with an altitude independent D/H profile, and with the D/H profiles generated in the case of the cold thermal profile and a water abundance of 10\,pr.\,\micron. In this calculation, we assume a whole-disk observation. For this optically thin line ($\tau\sim10^{-2}$) the emission mostly originates from the limb of the planet. We find that the HDO depletion induces a significant change in the line intensity. The open cloud model with a temperature-dependent fractionation coefficient decreases the contrast of the HDO-line by $\sim15\%$. The other three models decrease the contrast by 5--10\%. Such an effect is at the limit of the detectability of current observations.

A much more important effect arises if one exclusively probes atmospheric levels above the condensation level from orbiter observations in horizontal viewing. We simulate such limb observations in several cases spanning the different temperature profiles and water abundances. For this, we compare the HDO emission of two D/H profiles: a vertically independent D/H ratio, and that predicted by the closed cloud model with a temperature-dependent $\alpha$ (since it represents the average of our four cloud systems). The synthetic calculations are displayed on Fig.~5. They clearly show that the modeled D/H depletion leads to important differences in the line intensity from the constant D/H ratio case. The reason is that, one scale height above the condensation level, already 20 to 30\% of the deuterium is lost. Thus, probing at this altitude induces a 20 to 30\% decrease in the HDO line intensity. The  effect is even larger if the tangent altitude is higher.

There are current plans to fly a millimeter heterodyne instrument in Martian orbit, particularly in the framework of Martian micromissions. Our calculations show that if HDO lines are to be used to retrieve the vertical profile of \ho, the vapor pressure isotope effect needs to be considered. Conversely, comparing limb measurements on HDO and \ho\ (or ${\rm H_2{}^{18}O}$) is a promising strategy to detect and measure this effect.

\subsection{Neptune, Uranus and Titan: ${\rm\bf CH_3D}$}

\indent\indent The D/H ratio in Neptune has been measured from \chd\ both in the troposphere at 1.6\,\micron\ (de Bergh {\it et al.}\ 1990) and in the stratosphere at 8.6\,\micron\ (Orton {\it et al.}\ 1992, B\'ezard {\it et al.}\ 1997). The results of these three studies are compatible within their error bars. de Bergh {\it et al.}\ (1990) obtain $(9^{+12}_{-6})\times10^{-5}$ while Orton {\it et al.}\ (1992) give $(5.6^{+1.7}_{-1.3})\times10^{-5}$ and B\'ezard {\it et al.}\ (1997) $(5^{+4.4}_{-2.3})\times10^{-5}$. (Here, the values of de Bergh {\it et al.}\ (1990) and Orton {\it et al.}\ (1992) have been updated to account for the new value of the isotopic enrichment factor between \h\ and \ch\ recommended by Lecluse {\it et al.}\ (1996), $f=1.61\pm0.21$, cf.\ Feuchtgruber {\it et al.}\ (1999); we note that this value of $f$ is significantly different from that proposed by Fegley and Prinn (1988)). These three determinations can be affected by the VPIE, since the \chd-lines probe atmospheric levels well above the tropopause at 8.6\,\micron\ or around the condensation level at 1.6\,\micron, although the effect must be much smaller in the latter case.  In Uranus the only detection of \chd\ was obtained at 1.6\,\micron\ (de Bergh {\it et al.}\ 1986), yielding ${\rm D/H}=(5.4^{+7.0}_{-3.0})\times 10^{-5}$ for $f=1.68\pm 0.23$ (Lecluse {\it et al.}\ 1996, Feuchtgruber {\it et al.}\ 1999). Similarly to Neptune, the sounded atmospheric levels lie near the methane condensation level, and the D/H ratio can, in principle, be affected by isotopic fractionation.

Testing for this possibility, we model the \ch\ condensation process in Neptune using the temperature profile of Feuchtgruber {\it et al.}\ (1999) (Fig.~6a). We adopt the deep methane mole fraction of Baines {\it et al.}\ (1995), $f_{\rm CH_4}=0.022$. As for Mars, we assume that ice formation occurs at 100\% relative humidity. Figures 6b and 6c respectively display the fractionation coefficient $\alpha$ and the calculated methane vertical distribution as a function of pressure. It can be seen that methane condensation occurs at too shallow a level (1.5\,bar, 80\,K) to give rise to an inverse isotope effect. Thus, as for Mars, a loss of deuterium in the gas phase results is the net result, as shown on Fig.~6d.

According to Carlson {\it et al.}\ (1988) the rapid condensation of methane implies massive precipitation. In this case the open cloud system is the best model of the actual D/H depletion. Yet, the stratospheric measurement of methane, $f_{\rm CH_4}=(1.3^{+0.8}_{-0.7})\times10^{-3}$ (B\'ezard 1998), indicates a supersaturation with respect to the tropopause cold trap value ($1.3\times 10^{-4}$), implying that some ice particles are transported up to the stratosphere where they sublimate. We therefore adopt as a stratospheric estimate of the D/H depletion the value obtained at the pressure level, 0.7\,bar, where the methane mole fraction is equal to $1.3\times10^{-3}$. This yields $\delta_v=-1\%$ (Fig.~6d). Even if the observed supersaturation of \ch\ is not accounted for, the asymptotic (stratospheric) value of $\delta_v$ is only $-2.8\%$ (Fig.~6d). Thus the effect is too subtle to alter the interpretation of current measurements at 8.6\,\micron. This is naturally even more true for the 1.6\,\micron\ measurements. The same conclusion holds for Uranus, since its temperature profile and its methane abundance are similar to those of Neptune.

In Uranus and Neptune, the D/H ratio is also measured from HD (see Feuchtgruber {\it et al.}\ (1999)). However, comparing the measurements from HD with those from \chd\ cannot be used presently as an observational determination of the condensation fractionation effect, not only because the measurements are not accurate enough, but also because the fractionation factor between \ch\ and \h\ is not known with enough accuracy (Fegley and Prinn 1988, Lecluse {\it et al.}\ 1996).

In Titan there is no definite evidence for a permanent methane cloud, although Griffith {\it et al.}\ (1998) have reported observations of transient clouds near 15\,km. The methane tropospheric abundance is also not yet accurately determined. Many different methane profiles have been derived (Flasar 1983, Lellouch {\it et al.}\ 1989, Courtin {\it et al.}\ 1995). McKay {\it et al.}\ (1997) were even able to fit the Voyager occultation data with a constant \ch\ mixing ratio throughout the whole atmosphere. In this situation, we only calculate an upper limit of the VPIE in Titan.

For this, we assume that the fractionation coefficient is constant in the whole troposphere at its highest possible value, i.e.\ its value at the tropopause. Adopting the temperature profile of Yelle {\it et al.}\ (1997) with a tropopause at 71\,K, we find $\alpha=1.004$. Within this assumption Eq.~\ref{eq:1} can be solved analytically: $\delta_v=(1+\delta_0)(n_v/n_0)^{\alpha-1}-1$. Flasar (1983) pointed out that the surface humidity must be lower than 0.7 to keep the troposphere convectively stable, but here we adopt the conservative approach of a methane profile saturated throughout the troposphere. This again can only enhance the effect, and  yields a ratio of the mixing ratio at the surface to that at the tropopause of $n_v/n_0=0.21$. All these assumptions result in an upper limit of $\delta_v=-0.6\%$ to the D/H depletion in the stratosphere, totally negligible for the D/H determination from stratospheric methane. 

\section{Conclusions}

\indent\indent We have investigated the consequences of the vapor pressure isotopic effect on condensible deuterated species in planetary atmospheres. This effect induces a depletion of deuterium in the vapor phase above the condensation level. The magnitude of the depletion largely depends on the type of cloud system. In the atmospheres of Titan, Uranus and Neptune, the depletion in \chd\ is not detectable given current observational uncertainties, so that D/H ratios determinations are not altered by this isotopic fractionation.

In the atmosphere of Mars HDO is greatly affected. Depletions from 20\% to 99\% of the D/H ratio are possible a few scale heights above the condensation level. The magnitude of the effect depends primarily on the temperature profile. In particular the D/H measurement of Bjoraker {\it et al.}\ (1989), performed during a relatively warm period, does not suffer from isotopic fractionation. When the condensation level lies close to the surface, the isotopic fractionation effect induces variations of $\sim10\%$ in the HDO-line intensity for disk averaged millimetric observations. Such an effect might become detectable in the near future as the accuracy of these observations will improve.

The effect is much more spectacular in horizontal viewing. Limb sounding above the condensation level thus appears to be an efficient method to detect the HDO depletion. It is an interesting scientific objective for future Martian missions. Indeed, if detected and quantified, the vapor pressure isotope effect could enhance our understanding of the water condensation process and of the atmospheric circulation.
  
\newpage
\section*{Acknowledgements}

\indent\indent T.F.\ thanks N.\ Aghanim for her encouragements and Th.\ Encrenaz for a careful reading of the manuscript. We thank S.\ Joussaume for useful discussions on the physics of isotopic fractionation, and M.\ Smith and T.\ Owen for helpful reviews.

\newpage
\section*{References}

\noindent Armstrong, G.\ T.,
F.\ G.\ Brickwedde, and R.\ B.\ Scott 1953. The Vapor Pressures of the
Deuteromethanes. {\it J.\ Chem.\ Phys.}\ {\bf 21}, 1297--1298.\\

\noindent Baines, K.\ H., M.\ E.\ Mickelson,
L.\ E.\ Larson, and D.\  W.\ Ferguson 1995. The abundances of methane and
ortho/para hydrogen on Uranus and Neptune: Implications of New Laboratory 4--0
H$_2$ quadrupole line parameters. {\it Icarus} {\bf 114},  328--340.\\

\noindent Beer, R., and F.\ W.\ Taylor 1973.
The Abundance of CH$_3$D and the D/H Ratio in Jupiter. {\it Astrophys.\  J.}\
{\bf 179}, 309--328.\\

\noindent B\'ezard, B.\ 1998. Observations of
hydrocarbons in the giant planets with ISO-SWS. {\it Ann. Geophys.}\ {\bf 16},
C1037.\\

\noindent B\'ezard, B., T.\ Encrenaz,
H.\ Feuchtgruber, E.\ Lellouch, T.\ de Graauw, M.\ Griffin, and S.\ K.\ Atreya
1997. ISO--SWS Observations of Neptune. {\it Bull.\ Am.\ Astron.\ Soc.}\
{\bf 29}, 993.\\

\noindent Bigeleisen, J.,
C.\ B.\ Cragg, and M.\ Jeevanandam 1967. Vapor Pressures of Isotopic
Methanes---Evidence for Hindered Rotation. {\it J.\ Chem.\ Phys.}\ {\bf 47},
4335--4346.\\

\noindent Bjoraker, G.\ L., M.\ J.\ Mumma, and H.\ P.\ Larson 1989. Isotopic
abundance ratios for hydrogen and oxygen in the Martian atmosphere.
{\it Bull.\ Am.\ Astron.\ Soc.}\ {\bf 21}, 991.\\

\noindent Carlson, B.\ E., W.\ B.\ Rossow,
and G.\  S.\ Orton 1988. Cloud microphysics of the giant planets. {\it J.\
Atmos.\ Sci.}\ {\bf 45},  2066--2081.\\

\noindent Clancy, R.\ T., A.\ W.\ Grossman,
and D.\ O.\ Muhleman 1992. Mapping Mars water vapor with the Very Large Array.
{\it Icarus} {\bf 100}, 48--59.\\

\noindent Clancy, R. T., A. W. Grossman,
M. J. Wolff, P. B. James, D. J. Rudy, Y. N. Billawala, B. J. Sandor, S. W. Lee,
and D. O. Muhleman 1996. Water vapor saturation at low altitudes around Mars
aphelion: a key to Mars climate. {\it Icarus} {\bf 122}, 36--62.\\

\noindent Courtin, R., D.\ Gautier,
C.\ P.\ Mckay 1995. Titan's thermal emission spectrum: Reanalysis of the
Voyager infrared measurements. {\it Icarus} {\bf 114}, 144--162.\\

\noindent Dansgaard, W.\ 1964. Stable
isotopes in precipitation. {\it Tellus} {\bf 16}, 436--468.\\

\noindent de Bergh, C., B.\ L.\ Lutz,
T.\ Owen, J.\ Brault, and J.\ Chauville 1986. Monodeuterated methane in the
outer solar system. II. Its detection on Uranus at 1.6 microns.
{\it Astrophys.\ J.}\ {\bf  311}, 501--510.\\

\noindent de Bergh, C., B.\ L.\ Lutz,
T.\ Owen, and J.-P.\ Maillard 1990. Monodeuterated methane in the outer solar
system. IV. Its detection and abundance on Neptune. {\it Astrophys.\ J.}\ {\bf
355},  661--666.\\

\noindent Encrenaz, T., E.\ Lellouch,
J.\ Rosenqvist, P.\ Drossart, M.\ Combes, F.\ Billebaud, I.\ de Pater,
S.\ Gulkis, J.-P.\ Maillard, and G.\ Paubert 1991. The atmospheric composition
of Mars: ISM and ground-based observational data. {\it Ann.\ Geophys.}\
{\bf 9}, 797--803.\\

\noindent Fegley, B., and R.\ G.\ Prinn
1988. The predicted abundances of deuterium-bearing gases in the atmospheres of
Jupiter and Saturn. {\it Astrophys.\ J.}\ {\bf 326},  490--508.\\

\noindent Feuchtgruber, H.,
E.\ Lellouch, B.\ B\'ezard, T.\ Encrenaz, T.\ de Graauw, and  G.\ R.\ Davis
1997. Detection of HD in the atmospheres of Uranus and Neptune: a new 
determination of the D/H ratio. {\it Astron.\ Astrophys.}\ {\bf 341},
L17--L21.\\

\noindent Flasar, F.\ M.\ 1983. Oceans on Titan?
{\it Science} {\bf 221}, 55--57.\\

\noindent Griffith, C. A., T. Owen,
G. A. Miller, and T. Geballe 1998. Transient clouds in Titan's lower
atmosphere. {\it Nature} {\bf 395}, 575--578.\\

\noindent Hourdin, F., O.\ Talagrand,
R.\ Sadourny, R.\ Courtin, D.\ Gautier, and C.\ P.\ McKay 1995. Numerical
simulation of the general circulation of the atmosphere of Titan. {\it Icarus}
{\bf 117}, 358--374.\\

\noindent Jakosky, B.\ M.,
and R.\ M.\ Haberle 1992. The seasonal behavior of water on Mars. In
{\it Mars} (H.\ H.\ Kiefer, B.\ M.\ Jakosky, C.\ W.\ Snyder,
and M.\ S.\ Matthews, Eds.), pp.\ 969--1016. The Unviversity of Arizaona Press,
Tucson \& London.\\

\noindent Jancso, G., and A.\ Van Hook
1974. Condensed phase isotope effects (especially vapor pressure isotope
effects). {\it Chem.\ Rev.}\ {\bf 74}, 689--750.\\

\noindent Joussaume, S., R.\ Sadourny,
and J.\ Jouzel 1984. A general circulation model of water isotope cycles in
the atmosphere. {\it Nature} {\bf 311}, 24--29.\\

\noindent Jouzel, J.\ 1986. Isotopes in cloud
physics: multiphase and multistage condensation process. In {\it Handbook of
environmental isotope geochemistry, the terrestrial environment, Vol.~2},
(P.\ Fritz and J.\ C.\ Fontes, Eds.), pp.\ 61--112. Elsevier, New-York.\\

\noindent Jouzel, J., L.\ Merlivat,
and E.\ Roth 1975. Isotopic study of hail. {\it J.\ Geophys.\ Res.}\ {\bf 80},
5015--5030.\\

\noindent Jouzel, J., N.\ Brichet,
B.\ Thalmann, and B.\ Federer 1980. A numerical cloud model to interpret the
isotope content of hailstones. In {\it Proceedings of the Eighth Conference on
Cloud Physics, Clermont-Ferrand}. LAMP, Clermont-Ferrand.\\

\noindent Krasnopolsky, V.\ A.,
G.\ L.\ Bjoraker, M.\ J.\ Mumma, and D.\ E.\ Jennings 1997. High-resolution
spectroscopy of Mars at 3.7 and 8 $\mu$m: A sensitive search for H$_2$O$_2$,
H$_2$CO, HCl, and CH$_4$, and detection of HDO. {\it J.\ Geophys.\ Res.}\
{\bf 102}, 6525--6534.\\

\noindent Lecluse, C., F.\ Robert,
D.\ Gautier, and M.\ Guiraud 1996. Deuterium enrichment in giant planets.
{\it Planet.\ Space Sci.}\ {\bf  44}, 1579--1592.\\

\noindent Lellouch, E., A.\ Coustenis,
D.\ Gautier, F.\ Raulin, N.\ Dubouloz, and C.\ Fr\`ere 1989. Titan's
atmosphere and hypothesized ocean: A reanalysis of the Voyager 1
radio-occultation and Iris 7.7-micron data. {\it Icarus} {\bf 79}, 328--349.\\

\noindent Martin, T.\ Z.,
and H.\ H.\ Kieffer 1979. Thermal infrared properties of the Martian
atmosphere. II.\ The 15-micron band measurements. {\it J.\ Geophys.\ Res.}\
{\bf 84}, 2843--2852.\\

\noindent McKay, C.\ P., S.\ C.\ Martin,
C.\ A.\ Griffith, and R.\ M.\ Keller 1997. Temperature Lapse Rate and Methane
in Titan's Troposphere. {\it Icarus} {\bf  129}, 498--505.\\

\noindent Merlivat, L., and G.\ Nief
1967. Fractionnement isotopique lors des changements d'\'etat solide-vapeur et
liquide-vapeur de l'eau \`a des temp\'eratures inf\'erieures \`a 0$^{\circ}$C.
{\it Tellus} {\bf 19}, 122--127.\\

\noindent Orton, G.\ S., J.\ H.\ Lacy,
J.\ M.\ Achtermann, P.\ Parmar, and W.\ E.\ Blass 1992. Thermal spectroscopy
of Neptune: The stratospheric temperature, hydrocarbon abundances, and isotopic
ratios. {\it Icarus} {\bf 100}, 541--555.\\

\noindent Owen, T., J.-P. Maillard, C. de Bergh,
and B. L. Lutz 1988. Deuterium on Mars: the abundance of HDO and the value of
D/H. {\it Science} {\bf 240}, 1767--1770.\\

\noindent Owen, T.\ 1992. Deuterium in the solar
system. In {\it Astrochemistry of Cosmic Phenomena}, (P.\ Singh, Ed.),
pp.\ 97--101. Kluwer Academic, Dordrecht.\\

\noindent Read, P.\ L., M.\ Collins, F.\ Forget,
R.\ Fournier, F.\ Hourdin, S.\ R.\ Lewis, O.\ Talagrand, F.\ W.\ Taylor,
and N.\ P.\ J.\ Thomas 1997. A GCM climate database for mars: for mission
planning and for scientific studies. {\it Adv.\ Space Res.}\ {\bf 19},
1213--1222.\\

\noindent Taylor, C.\ B.\ 1972. The vertical
variations of the isotopic concentrations of tropospheric water vapour over
continental Europe and their relationship to tropospheric structure.
{\it N.Z.\ Dep.\ Sci.\ Ind.\ Res., Inst.\ Nucl.\ Sci.\ Rep.},
INS-R-107, 45 pp.\\

\noindent Yelle, R.\ V., D.\ F.\ Strobel,
E.\ Lellouch, and D.\ Gautier 1997. In {\it Huygens: Science, Payload and
Mission} (A.\ Wilson, Ed.), pp.\ 253-256. ESA SP-1177, European Space Agency,
Noordwijk.\\


\newpage
\noindent {\bf Table I}: Loss of HDO in Mars' atmosphere: difference between the VPIE-fractionated column densities and the non-fractionated column density.\\[1cm]
\begin{tabular}{cccccccc} 
\hline
\multicolumn{1}{c}{}& \multicolumn{1}{c}{Water}& \multicolumn{3}{c}{Open Model}& \multicolumn{3}{c}{Closed Model}\\
{}& abundance (pr.\,\micron)& Cold& Mean& Hot& Cold& Mean& Hot\\ \hline
Constant& 3& 1.3\%& 0.19\%& $<0.1\%$& 0.81\%& 0.11\%& $<0.1\%$\\
$\alpha$& 10& 3.1\%& 0.33\%& $<0.1\%$& 1.9\%& 0.20\%& $<0.1\%$\\
& 30& 10\%& 0.65\%& $<0.1\%$& 6.4\%& 0.40\%& $<0.1\%$\\ \hline
Temperature& 3& 2.5\%& 0.38\%& $<0.1\%$& 1.6\%& 0.25\%& $<0.1\%$\\
dependent& 10& 5.1\%& 0.63\%& $<0.1\%$& 3.4\%& 0.41\%& $<0.1\%$\\
$\alpha$& 30& 15\%& 1.1\%& $<0.1\%$& 10\%& 0.73\%& $<0.1\%$\\ \hline
\end{tabular}

\newpage
\section*{Figure Captions}

\noindent {\bf Figure 1} D/H depletion in Mars' atmosphere for the cold temperature profile. Panel a: Temperature profile. Panel b: $\delta_v$ vertical profiles calculated for a water abundance of 3 pr.\,\micron\ with an open cloud model for temperature-dependent $\alpha$ (OM1, solid thick line) and for a constant $\alpha$ (OM2, dashed thick line), with a closed cloud model for temperature-dependent $\alpha$ (CM1, solid thin line) and for a constant $\alpha$ (CM2, dashed thin line). Panels c and d: Same for 10 and 30 pr.\,\micron.\\

\noindent {\bf Figure 2} Same as figure 1 for the mean temperature profile.\\

\noindent {\bf Figure 3} Same as figure 1 for the hot temperature profile.\\

\noindent {\bf Figure 4} Comparison of disk averaged 226-GHz HDO line in Mars' atmosphere for a nonfractionated profile (tripledotted-dashed line) with that obtained for the four different HDO profiles calculated with the cold profile and 10 pr.\,\micron (see Fig.~1c): OM1 (solid thick line), OM2 (dashed thick line), CM1 (solid thin line) and CM2 (dashed thin line).\\

\noindent {\bf Figure 5} Comparison of limb observations of the 226-GHz HDO line calculated with a non-fractionated profile (solid line) with that calculated with a closed cloud model and a temperature-dependent $\alpha$ (CM1, dashed-dotted line) in four different cases. Panel a: Cold temperature profile, water abundance of 30 pr.\,\micron\ and a tangent altitude of 5\,km. Panel b: Cold temperature profile, water abundance of 10 pr.\,\micron\ and a tangent altitude of 20\,km. Panel c: Mean temperature profile, water abundance of 10 pr.\,\micron\ and a tangent altitude of 30\,km. Panel d: Cold temperature profile, water abundance of 3 pr.\,\micron\ and a tangent altitude of 60\,km.\\
 
\noindent {\bf Figure 6} D/H depletion in Neptune's atmosphere. Panel a: Temperature profile. Panel b: Fractionation coefficient as a function of pressure. Panel c: \ch\ vertical profile. Panel d: D/H vertical profile calculated with an open cloud model (solid line) and a closed cloud model (dashed line).\\

\newpage
\pagestyle{empty}
\begin{figure}[!h]
  \begin{center}
  \leavevmode
  \centerline{\epsfig{file=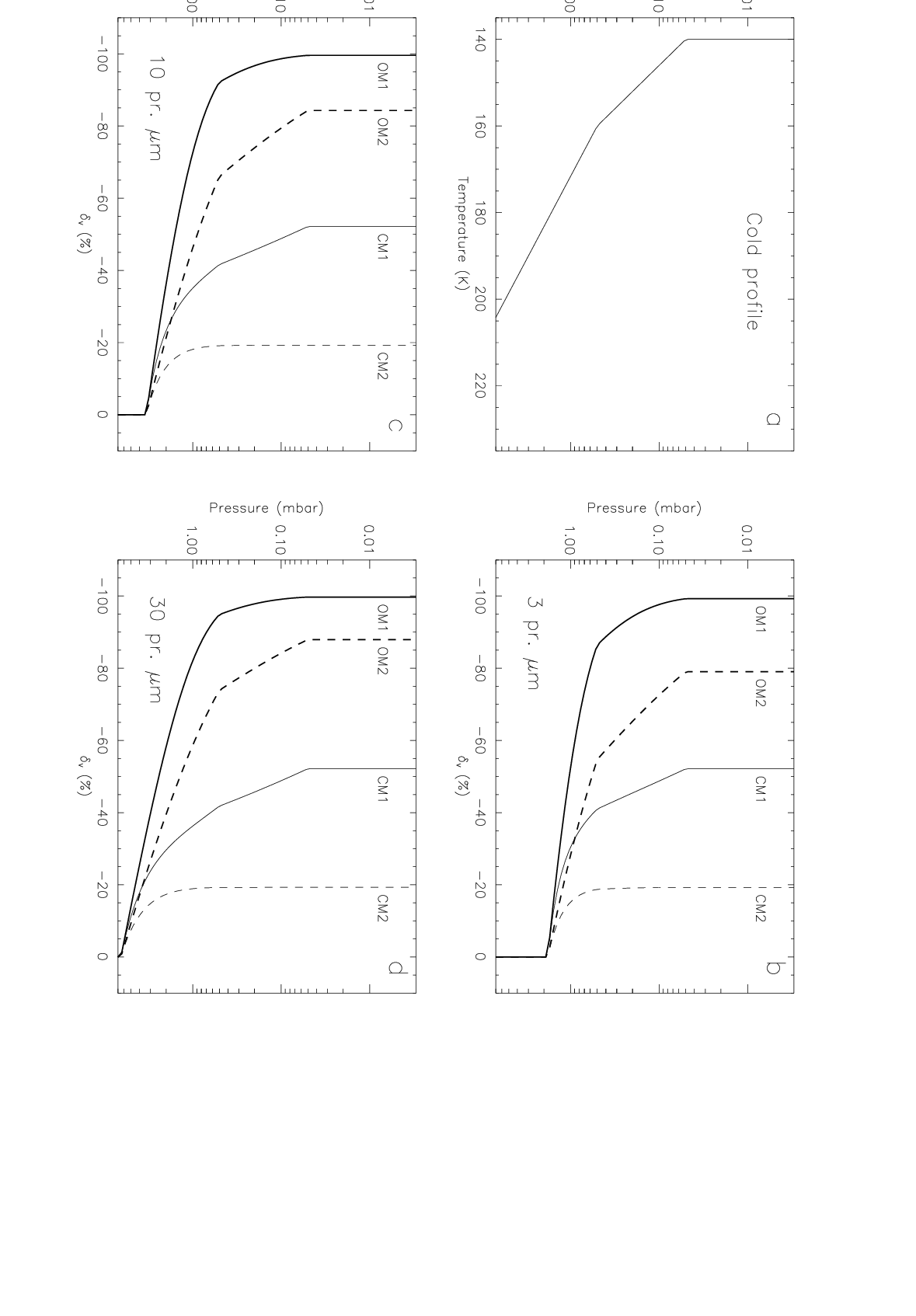}}
  \end{center}
  \caption{Fouchet, Figure~1}
\end{figure}

\newpage
\pagestyle{empty}
\begin{figure}[!h]
  \begin{center}
  \leavevmode
  \centerline{\epsfig{file=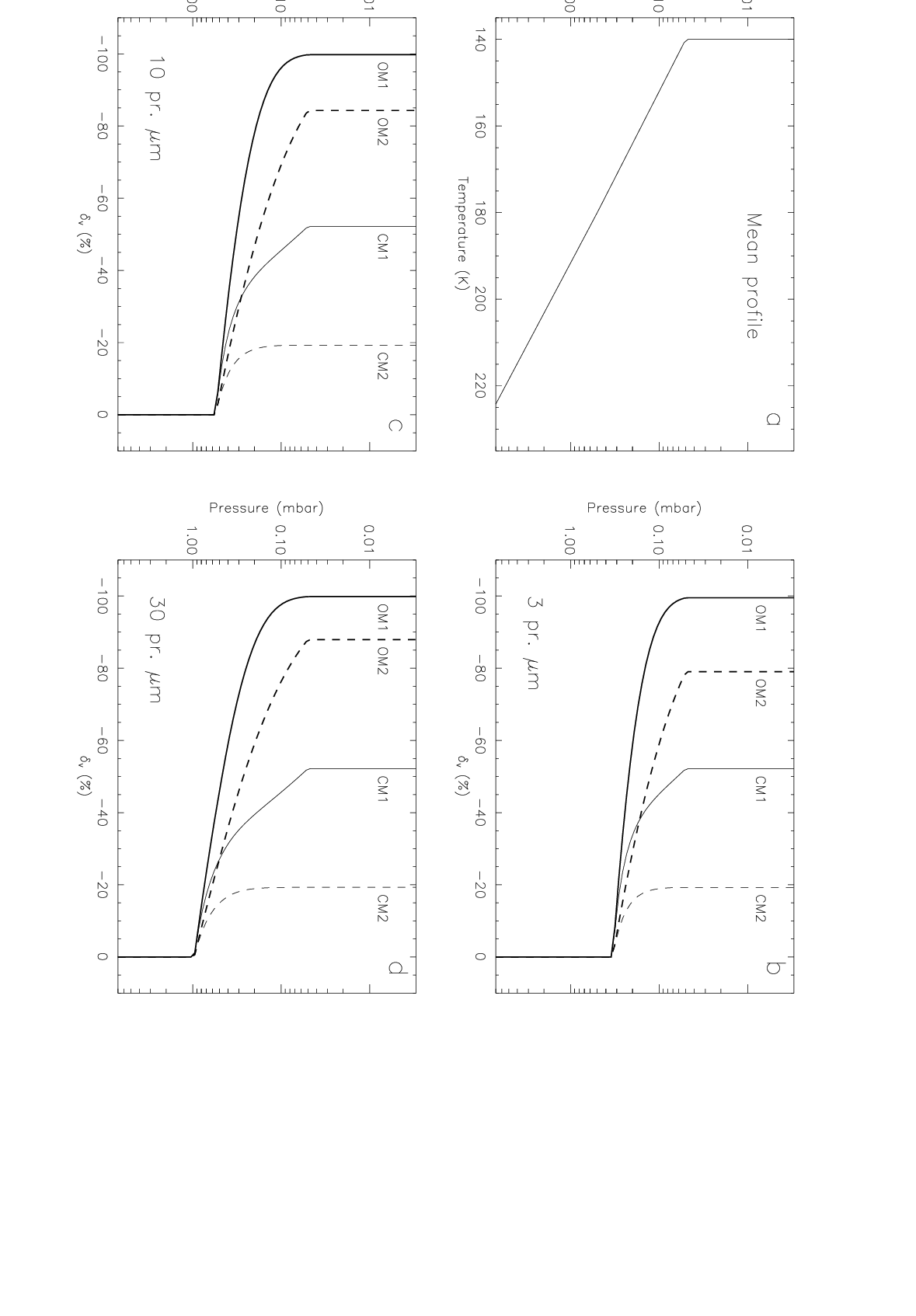}}
  \end{center}
\end{figure}

\newpage
\pagestyle{empty}
\begin{figure}[!h]
  \begin{center}
  \leavevmode
  \centerline{\epsfig{file=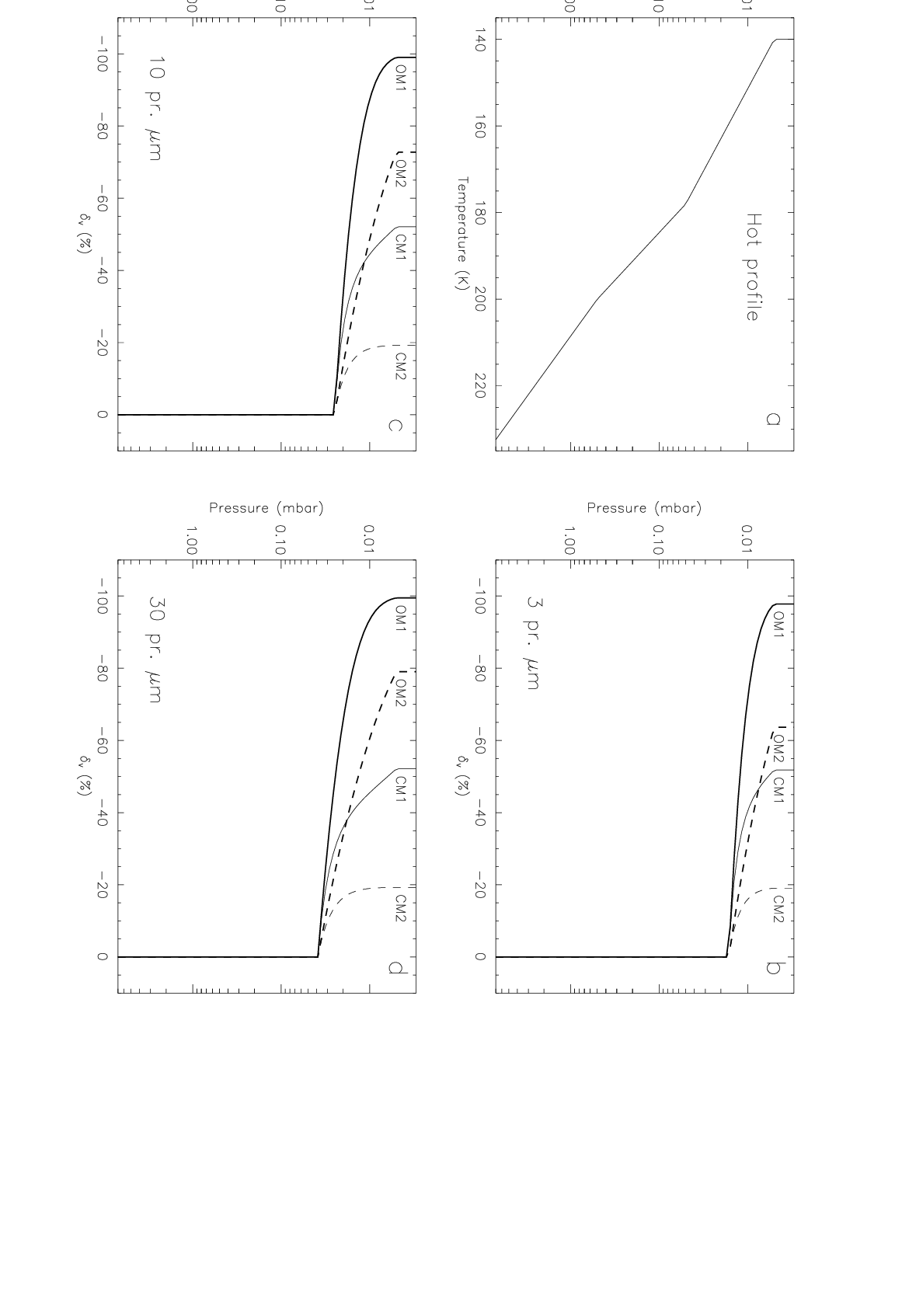}}
  \end{center}
\end{figure}

\newpage
\pagestyle{empty}
\begin{figure}[!h]
  \begin{center}
  \leavevmode
  \centerline{\epsfig{file=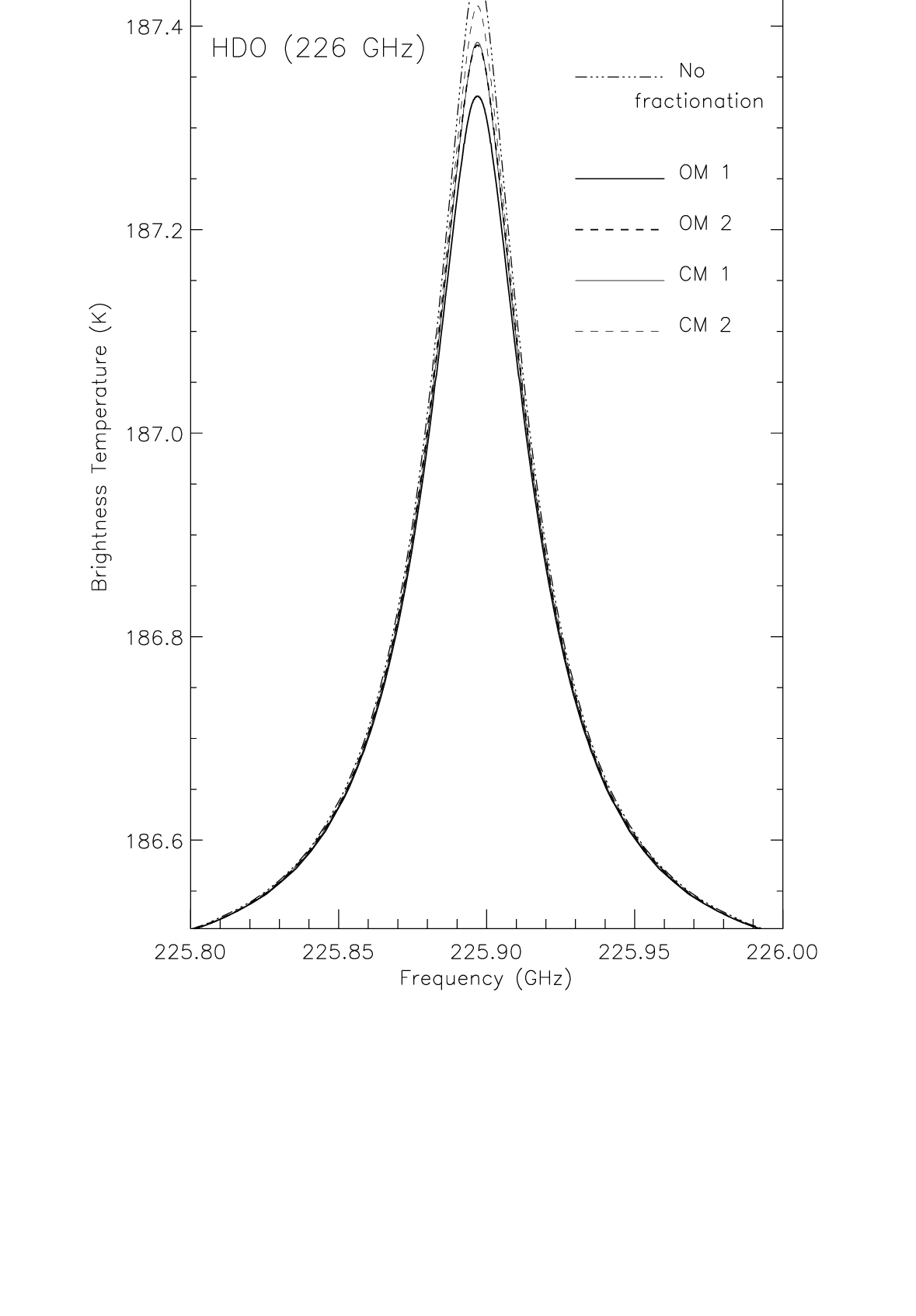}}
  \end{center}
\end{figure}

\newpage
\pagestyle{empty}
\begin{figure}[!h]
  \begin{center}
  \leavevmode
  \centerline{\epsfig{file=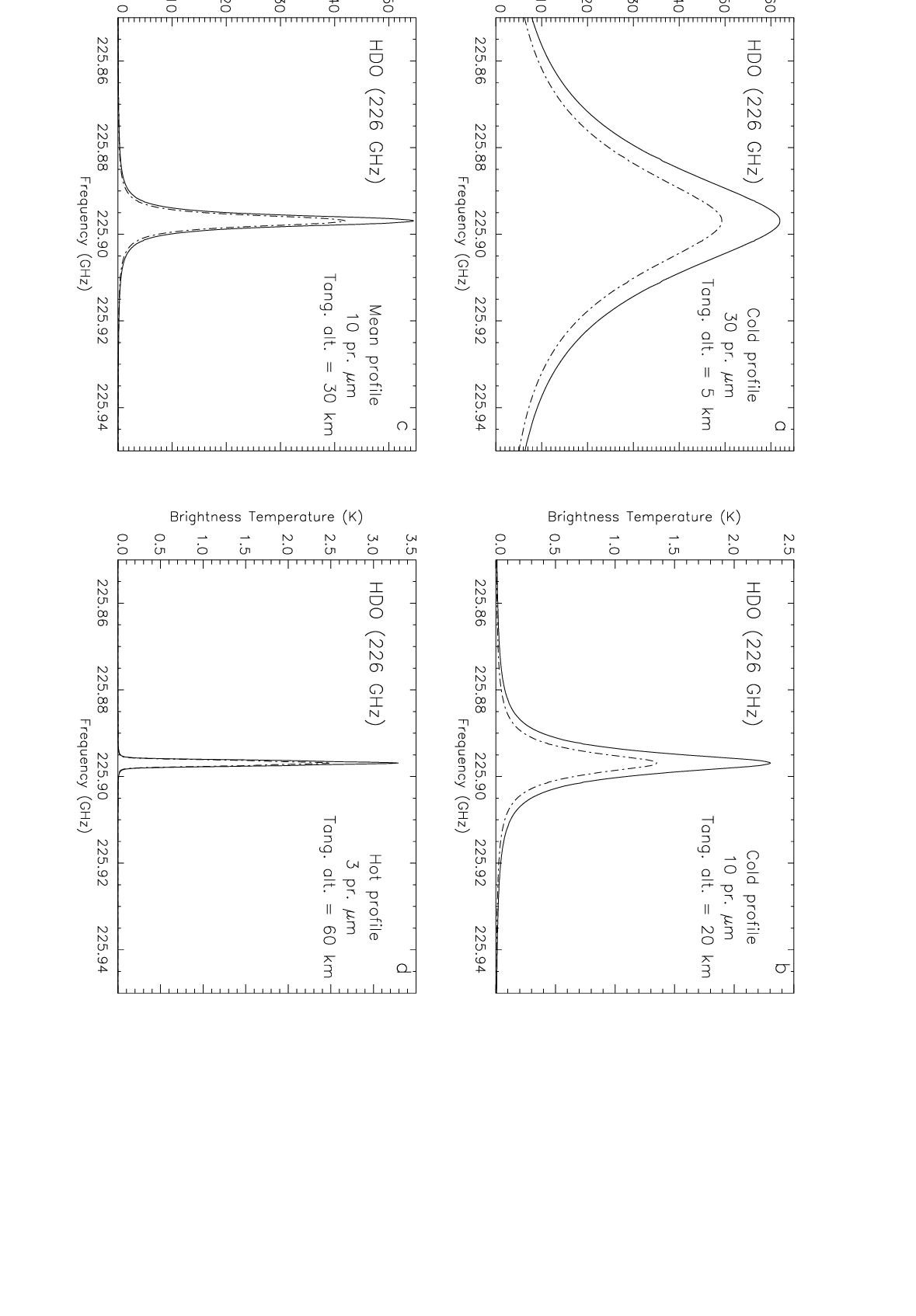}}
  \end{center}
\end{figure}

\newpage
\pagestyle{empty}
\begin{figure}[!h]
  \begin{center}
  \leavevmode
  \centerline{\epsfig{file=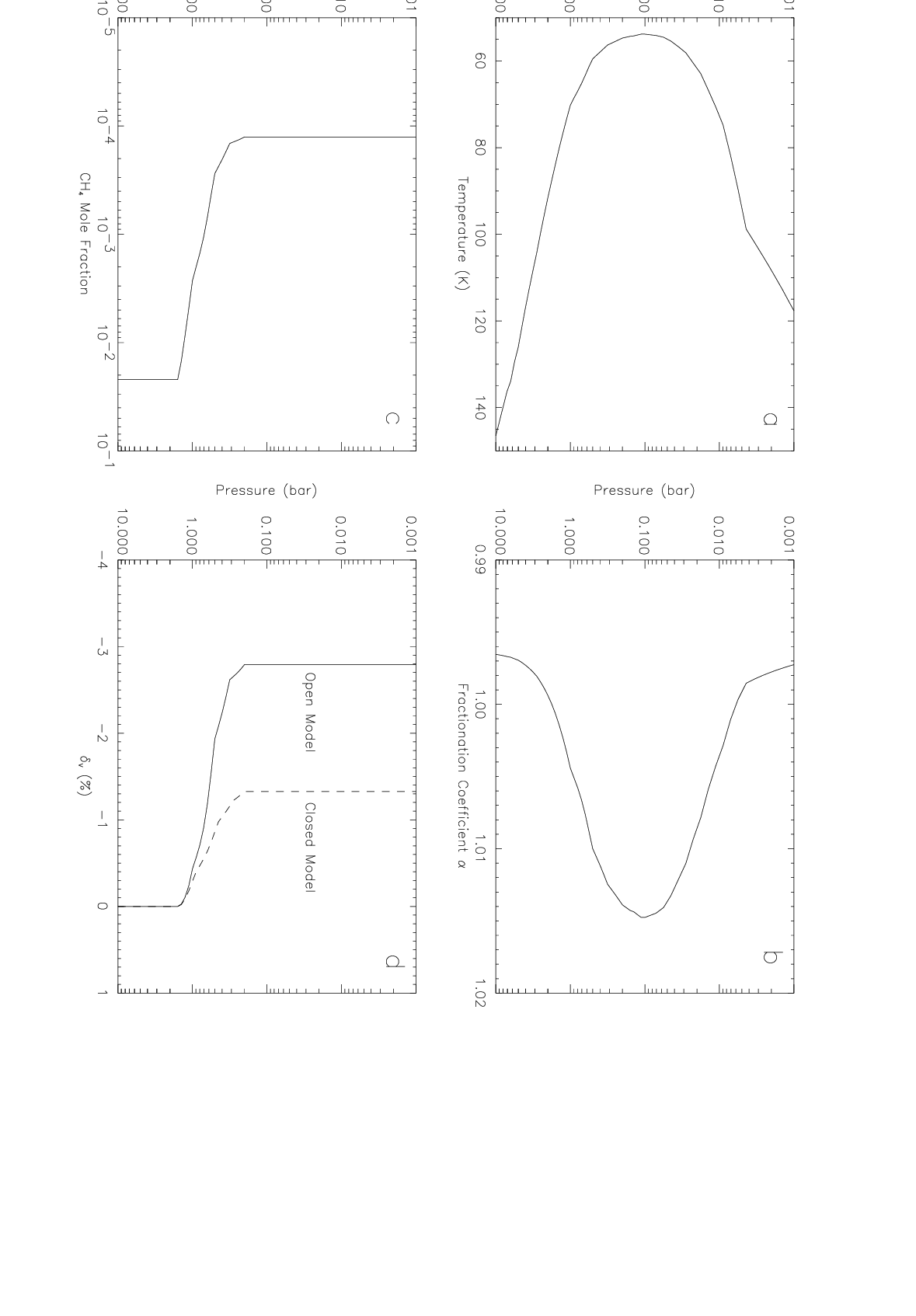}}
  \end{center}
\end{figure}

\end{document}